\documentclass[12pt]{article}
\usepackage[dvips]{graphicx}

\usepackage{graphics,epsfig}

\renewcommand{\bar}[1]{\overline{#1}}

\begin{document}

\begin{center}
{\Large
\bf Decay Constants of $D_{sJ}^*$(2317) and $D_{sJ}$(2460)}\\


\vspace{1.5cm}

Dae Sung Hwang${}^{(a)}$ and Do-Won Kim${}^{(b)}$\\
{\it{a: Department of Physics, Sejong University, Seoul 143--747,
Korea}}\\
{\it{b: Department of Physics, Kangnung National University,
Kangnung 210-702, Korea}}\\

\vspace{2.5cm}

{\bf Abstract}\\
\end{center}
\noindent
The resonances $D_{sJ}^*$(2317) and $D_{sJ}$(2460) which are considered to be the
$(0^+,1^+)$ doublet composed of charm and strange quarks have been discovered recently.
Using the method of Rosner which is based on the factorization hypothesis,
we calculate the lower bounds of the decay constants of
these states from the branching ratios of $B\to D D_{sJ}$ measured by Belle and Babar.
Our result shows that the decay constant of $D_{sJ}$(2460) is about twice 
that of $D_{sJ}^*$(2317) on the contrary to the naive expectation of the heavy quark
symmetry which gives their equality.
We show that this big deviation originates from the large internal motion of
quarks inside these $P$-wave states and that our result is in good accord with the
relativistic quark model calculation.
\\

\vfill


\noindent
$^a$e-mail: dshwang@sejong.ac.kr\\
$^b$e-mail: dwkim@kangnung.ac.kr\\
\thispagestyle{empty}
\pagebreak

\setlength{\baselineskip}{13pt}

\section{Introduction}

The resonances $D_{sJ}^*$(2317) and $D_{sJ}$(2460) composed of charm and strange quarks
have been discovered recently by
the BaBar \cite{babar03}, CLEO \cite{cleo03}, and Belle \cite{belle03} Collaborations.
Their decay patterns suggest that they are $0^+$ and $1^+$ states, respectively,
in the quark-model classification.
Bardeen et al. \cite{BEH} considered these states to be the
$(0^+,1^+)$ doublet which has $j=1/2$ of the light degree of freedom and studied them
with effective Lagrangians based on the chiral symmetry
in heavy-light meson systems.

The measured mass of $D_{sJ}^*$(2317),
2317.4$\pm$0.9 MeV \cite{pdg04} which is 40.9$\pm$1.0
MeV below the threshold of $D^0K^+$, was considered surprisingly low compared to
the predictions of the potential model calculations.
For example, the prediction of the $1{}^3P_0$ mass by
Isgur and Godfrey \cite{IG} was 2.48 GeV,
and that by Eichten and Di Pierro \cite{EP} was 2.487 GeV,
which are about 160 and 170 MeV higher than the measured mass of $D_{sJ}^*(2317)$.
There have been many theoretical investigations which aimed to explain the measured
low mass of $D_{sJ}^*(2317)$
\cite{CJ,BCL,BR,ChengHou,Szcz,BPP,HwangKim,SimonovTjon}.
For example,
Barnes et al. \cite{BCL} considered a mixing between two molecular states
$|D^0K^+>$ and $|D^+K^0>$
and pointed out the importance of a very strong coupling between the $c\bar{s}$ bound
and $DK$ continuum states, as required to induce binding.
Van Beveren and Rupp \cite{BR} described $D_{sJ}^*(2317)$ as a quasibound scalar
$c\bar{s}$ state in a unitarized meson model, owing its existence to the strong
coupling to the nearby $S$-wave $DK$ threshold.
Browder et al. \cite{BPP} proposed a mixing between the $q{\bar{q}}$ and 4-quark states
and assigned a linear combination with less mass as $D_{sJ}^*(2317)$.
Ref. \cite{HwangKim} calculated the mass shift of $D_{sJ}^*(2317)$ quantitatively
by using the coupled channel effect and could explain naturally the observed mass.

Bell \cite{belle03} and Babar \cite{babar04} measured the branching ratios of
the exclusive modes
$B\to D D_{sJ}^*(2317)[D_s^+\pi^0]$,
$B\to D D_{sJ}(2460)[D_s^{*+}\pi^0]$
and $B\to D D_{sJ}(2460)[D_s^{+}\gamma ]$.
Rosner calculated the decay constant of $D^{-}_s$ meson by
relating the the differential distributions
$d\Gamma ({\overline{B}}^0\to D^{(*)+}l^-{\overline{\nu}}_l)/dq^2$
and the rates the color-favored decays ${\overline{B}}^0\to D^{(*)+}D^{-}_s$
under the factorization hypothesis
\cite{Rosner90,Rosner01}.
Using the method of Rosner, we calculate the lower bounds of the decay constants of
$D_{sJ}^*$(2317) and $D_{sJ}$(2460) from
the partial branching ratios of $B\to D D_{sJ}$ measured by Bell and Babar.
Our result shows that the decay constant of $D_{sJ}$(2460) is about twice 
that of $D_{sJ}^*$(2317) on the contrary to the expectation of the heavy quark
symmetry which gives their equality.
We show that this big deviation originates from the large internal motion of
quarks inside these $P$-wave states and that our result is in good accord with the
relativistic quark model calculation.

In section 2.1 we calculate the lower bound of the decay constants of
$D_{sJ}^*$(2317) and $D_{sJ}$(2460), and estimate the ratio of these
decay constants. In section 2.2 we compare our results with the results of
the relativistic quart model calculation by Veseli and Dunietz.
Section 3 is conclusion, in which we discuss the physical implications of our results. 


\section{Decay Constants of $D_{sJ}^*$(2317) and $D_{sJ}$(2460)}

\subsection{Extraction from Measured Branching Ratios of $B\to D D_{sJ}$}

{}From Lorentz invariance one finds the decomposition of the
hadronic matrix element in terms of hadronic form factors:
\begin{eqnarray}
<D^+(p_D )|J_\mu |{\bar{B}}^0(p_B)>
&=&
\left[ (p_B+p_D )_\mu
-{m_B^2-m_D^2\over q^2}q_\mu \right] \, F_1^{BD}(q^2)
\nonumber \\
\cr
&&
+{m_B^2-m_D^2\over q^2}\, q_\mu \, F_0^{BD}(q^2),
\label{a1}
\end{eqnarray}
where $J_\mu = {\bar{c}}\gamma_\mu b$ and
$q_\mu =(p_B-p_D )_\mu$.
In the rest frame of the decay products, $F_1^{BD}(q^2)$ and $F_0^{BD}(q^2)$
correspond to $1^-$ and $0^+$ exchanges, respectively.
At $q^2=0$ we have the constraint
$F_1^{BD}(0)=F_0^{BD}(0)$
since the hadronic matrix element in (\ref{a1}) is nonsingular
at this kinematic point.

When the lepton masss is ignored,
the $q^2$ distribution of the semi-leptonic
decay rate, in the allowed range $0\le q^2\le (m_B-m_D)^2$, is given by
\begin{equation}
{d\Gamma ({\overline{B}}^0\to D^+l^-{\overline{\nu}}_l)\over dq^2}=
{G_F^2\over 24\pi^3}|V_{cb}|^2(K(q^2))^3|F_1^{BD}(q^2)|^2 \ ,
\label{semilep1}
\end{equation} 
where $K(q^2)=((m_B^2+m_D^2-q^2)^2-4m_B^2m_D^2)^{1/2}/2m_B$.

In the factorization hypothesis the effective Hamiltonian
${\cal H}_{\rm eff}$ for the process $B\to D D_{sJ}$
is written as \cite{wsb}
\begin{equation}
{\cal H}_{\rm eff}={G_F\over {\sqrt{2}}}V_{cb}V_{cs}^*
\Bigl( a_1[{\bar{s}}\Gamma^\mu c]_H
[{\bar{c}}\Gamma_\mu b]_H
+a_2[{\bar{c}}\Gamma^\mu c]_H
[{\bar{s}}\Gamma_\mu b]_H\Bigr)
\ +\ {\rm H.C.},
\label{b4}
\end{equation}
where $\Gamma^\mu = \gamma^\mu (1-\gamma_5)$ and
the subscript $H$ stands for $hadronic$ implying that the
Dirac bilinears inside the brackets be treated as interpolating
fields for the mesons and no further Fierz-reordering need be done.
The QCD corrections $a_1$ and $a_2$ have the values
$a_1\sim 1$ and $a_2\sim 0.25$ \cite{Fleischer}.
Luo and Rosner used $|a_1|=1.05$ in their calculation \cite{Rosner01}.

For the two body hadronic decay, in the rest frame of initial meson
the differential decay rate is given by
\begin{equation}
d\Gamma ={1\over 32\pi^2}|{\cal M}|^2
{|{\bf p}_1|\over M^2}d\Omega ,
\label{b7}
\end{equation}
\begin{equation}
|{\bf p}_1|=
{[(M^2-(m_1+m_2)^2)(M^2-(m_1-m_2)^2)]^{{1/2}}\over 2M},
\label{b8}
\end{equation}
where $M$ is the mass of initial meson, and $m_1$ ($m_2$) and
${\bf p}_1$ are the mass and momentum of one of final mesons.
By using (\ref{a1}), (\ref{b4}),
$<0|\Gamma_\mu |D_{s0}^{*}(q)>=iq_\mu f_{D_{s0}^{*}}$ and
$<0|\Gamma_\mu |D_{s1}^{*}(q,\varepsilon )>
=\varepsilon_\mu (q) m_{D_{s1}^{*}} f_{D_{s1}^{*}}$,
(\ref{b7}) gives the following formulas for the branching ratios of the
process ${\bar{B}}^0\rightarrow D^+ D_{s0}^{*-}$ and
${\bar{B}}^0\rightarrow D^+ D_{s1}^{*-}$:
\begin{eqnarray}
{\cal{B}} ({\bar{B}}^0\rightarrow D^+ D_{s0}^{*-})
&=&\left({G_Fm_B^2\over {\sqrt{2}}}\right)^2\,
|V_{cs}|^2\, {1\over 16 \pi }\, {m_B\over {\Gamma}_B}\, |a_1|^2\,
{f_{D_{s0}^{*}}^2 \over m_B^2}\, |V_{cb}\, F_0^{BD}(m_{D_{s0}^{*}}^2)|^2
\nonumber\\
& &\hspace*{-1.5cm} \times \Bigl( 1-{m_D^2\over m_B^2}{\Bigr)}^2\,
\left[\Bigl( 1-({m_D+m_{D_{s0}^{*}}\over m_B})^2{\Bigr)}
\Bigl( 1-({m_D-m_{D_{s0}^{*}}\over m_B})^2{\Bigr)}\right]^{1/2} ,
\qquad
\label{b11}\\
{\cal{B}} ({\bar{B}}^0\rightarrow D^+ D_{s1}^{*-})
&=& \left({G_Fm_B^2\over {\sqrt{2}}}\right)^2\,
|V_{cs}|^2\, {1\over 16 \pi }\, {m_B\over {\Gamma}_B}\,
|a_1|^2\,
{f_{D_{s1}^{*}}^2 \over m_B^2}\, |V_{cb}\, F_1^{BD}(m_{D_{s1}^{*}}^2)|^2
\nonumber\\
&\times &
\left[\Bigl( 1-({m_D+m_{D_{s1}^{*}}\over m_B})^2{\Bigr)}
 \Bigl( 1-({m_D-m_{D_{s1}^{*}}\over m_B})^2{\Bigr)}
\right]^{3/2}
\ .\qquad
\label{b12}
\end{eqnarray}

For the $B$ to $D$ meson (heavy to heavy) transition form factors, the heavy
quark effective theory gives \cite{IW8990}
\begin{equation}
F_1(q^2)={m_B+m_D\over 2{\sqrt{m_B m_D}}}\, {\cal G}(\omega)\ , \qquad
F_0(q^2)={2{\sqrt{m_B m_D}}\over m_B+m_D}\, {\omega + 1 \over 2}\,
{\cal G}(\omega)\ ,
\label{ff1}
\end{equation}
where $\omega = (m_B^2 + m_D^2 - q^2)/(2m_B m_D) = E_D/m_D$
($E_D$ is the energy of $D$ meson in the $B$ meson rest frame), and
${\cal G}(\omega)$ is a form factor which becomes the Isgur-Wise function in
the infinite heavy quark mass limit.
We use the parameterization of ${\cal G}(\omega)$
given in \cite{Caprini,Battaglia},
\begin{equation}
{{\cal G}(\omega)\over {\cal G}(1)} \approx
1-8{\rho}_{\cal G}^2 z + (51. {\rho}_{\cal G}^2 - 10. )z^2 -
(252. {\rho}_{\cal G}^2 - 84. ) z^3\ ,
\label{ff2}
\end{equation}
with
\begin{equation}
z = {{\sqrt{\omega +1}} - {\sqrt{2}} \over {\sqrt{\omega +1}} + {\sqrt{2}}}\ .
\label{ff3}
\end{equation}
We use the world average values given in \cite{Battaglia},
\begin{equation}
{\cal G}(1)|V_{cb}|\times 10^3=41.3 \pm 2.9 \pm 2.7\ , \qquad
{\rho}_{\cal G}^2=1.19 \pm 0.15 \pm 0.12 \ .
\label{ff4}
\end{equation}
The errors in (\ref{ff4}) give the error of ${\cal G}(\omega)$
by 12 \% for $D_{s0}^*$(2317) and by 11 \% for $D_{s1}'$(2460),
and they reduce to the same amounts of the errors for $f_{D_{s0}^*}$
and $f_{D_{s1}'}$, respectively, since we calculate these decay constants
by using Eqs. (\ref{b11}) and (\ref{b12}).
However, these errors are almost cancelled in the ratio
$f_{D_{s1}'}/f_{D_{s0}^*}$.

\begingroup
\begin{table}[t]
\label{tab:eigenvalue}
\begin{center}
\begin{tabular}{|l|l|c|c|c|}
\hline\hline
 & & & &
\\
\ \ \ Group \ \ \ & \ \ \ Decay Mode \ \ \  &
 $|a_1|\, f_{D_{s1}'}$ (MeV) & $|a_1|\, f_{D_{s0}^*}$ (MeV) &
\ \ \ \ $f_{D_{s1}'}/f_{D_{s0}^*}$\ \ \ \ \\
 & & & &
\\
\hline\hline
   &
\ \ \ $B^0\rightarrow D^- D_{s1}'^+$ \ \ \ &
175 $\pm$ 39 & & 2.61 $\pm$ 0.89 \\
\ \ \ Belle\ \ \ &
\ \ \ $B^0\rightarrow D^- D_{s0}^{*+}$ \ \ \ & &
\ 67 $\pm$ 20 & \\
\cline{2-5}
   &
\ \ \ $B^+\rightarrow {\overline D}^0 D_{s1}'^+$ \ \ \ &
126 $\pm$ 33 & & 2.00 $\pm$ 0.72 \\
   &
\ \ \ $B^+\rightarrow {\overline D}^0 D_{s0}^{*+}$ \ \ \ & &
\ 63 $\pm$ 19 & \\
\hline\hline
   &
\ \ \ $B^0\rightarrow D^- D_{s1}'^+$ \ \ \ &
189 $\pm$ 47 & & 1.95 $\pm$ 0.64 \\
\ \ \ Babar\ \ \ &
\ \ \ $B^0\rightarrow D^- D_{s0}^{*+}$ \ \ \ & &
\ 97 $\pm$ 27 & \\
\cline{2-5}
   &
\ \ \ $B^+\rightarrow {\overline D}^0 D_{s1}'^+$ \ \ \ &
173 $\pm$ 43 & & 2.47 $\pm$ 0.91 \\
   &
\ \ \ $B^+\rightarrow {\overline D}^0 D_{s0}^{*+}$ \ \ \ & &
\ 70 $\pm$ 22 & \\
\hline\hline
\multicolumn{2}{|c|}{Average} & 166 $\pm$ 20 & \ 74 $\pm$ 11 &
2.26 $\pm$ 0.41 \\
\hline\hline
\end{tabular}
\end{center}
\vspace*{-0.5cm}
\caption{The results for the lower bounds of the decay constants of
$D_{s1}'^+$(2460) and $D_{s0}^{*+}$(2317) and their ratio.
The values in the third column were obtained from the sum of the branching ratios
$B\to D D_{sJ}(2460)[D_s^{*+}\pi^0]$
and $B\to D D_{sJ}(2460)[D_s^{+}\gamma ]$,
and those in the fourth column from the branching ratio
$B\to D D_{sJ}^*(2317)[D_s^+\pi^0]$ measured by Belle \cite{belle03} and
Babar \cite{babar04}.
The values in the fifth column are the ratios of the values in the third and
fourth columns.}
\end{table}
\endgroup

We extract the lower bounds of
the the decay constants of $D_{s0}^*$(2317) and $D_{s1}'$(2460)
from Eqs. (\ref{b11}) and (\ref{b12}) by using the branching ratios
$B\to D D_{sJ}^*(2317)[D_s^+\pi^0]$,
$B\to D D_{sJ}(2460)[D_s^{*+}\pi^0]$
and $B\to D D_{sJ}(2460)[D_s^{+}\gamma ]$
measured by
Belle \cite{belle03} and Babar \cite{babar04},
and the above form factor ${\cal G}(\omega)$.
The results are presented in Table 1.
The value in the fifth column in Table 1 is the ratio of the lower bounds of the decay
constants given in the third and fourth columns. However, even in the situation that
the experimental values of
the branching ratios $B\to D D_{sJ}(2460)$ and $B\to D D_{sJ}^*(2317)$ are raised by
other partial branching ratios in addition to those considered here, it is expected
that the value in the fifth column does not change much because of the cancellation
in the ratio. Therefore, we expect that the value in the fifth column is
close to the ratio of the decay constants themselves $f_{D_{s1}'}$ and $f_{D_{s0}^*}$.

\subsection{Comparison with Relativistic Quark Model Calculation}

When we take the internal motion of quarks inside a meson into account,
the decay constants
of the $S$-wave pseudo-scalar ($J^P_j=0^-_{1/2}$) and vector ($1^-_{1/2}$) mesons,
where the subscript $j$ stands for the angular momentum of the light degree of
freedom in the $j$-$j$ coupling scheme of the heavy(${\bar{Q}}$)-light($q$) meson,
are given by \cite{HK}
\begin{equation}
f_i={2{\sqrt{3}}\over {\sqrt{M}}}{\sqrt{4\pi}}\int_0^{\infty}
{p^2dp\over (2\pi)^{3/2}}
{\sqrt{(m_q+E_q)(m_{\bar{Q}}+E_{\bar{Q}})\over 4E_qE_{\bar{Q}}}}\, F_i(p)\ ,
\label{ff5}
\end{equation}
with
\begin{eqnarray}
F_{0^-_{1/2}}(p)&=&\Big[ 1 - {p^2 \over (m_q+E_q)(m_{\bar{Q}}+E_{\bar{Q}})} \Big]
\, R_{n0}(p)\ , \ \ \
\nonumber\\
F_{1^-_{1/2}}(p)&=& \Big[ 1 + {1\over 3} {p^2 \over (m_q+E_q)(m_{\bar{Q}}+E_{\bar{Q}})} \Big]
\, R_{n0}(p) \ .
\label{ff6}
\end{eqnarray}
In the limit $m_{\bar{Q}}\to \infty$, from (\ref{ff5}) and (\ref{ff6})
both $f_{0^-_{1/2}}$ and $f_{1^-_{1/2}}$ become ${\sqrt{12/M}}\, |\psi (0)|$,
which is the Van Royen-Weisskopf formula \cite{RW}.
However, since in the $D_s$ meson system there is an appreciable contribution
of the internal motion of quarks to the decay constants given by (\ref{ff5})
and (\ref{ff6}), $f_{1^-_{1/2}}$ becomes larger than $f_{0^-_{1/2}}$.
Ref. \cite{HK} obtained the results:
$f_{D_s}=309$ MeV, $f_{D_s^*}=362$ MeV, and $f_{D_s^*}/f_{D_s} = 1.17$ by
averaging the values obtained from six different potential models.
For reference, the results of Ref. \cite{HK} for $B_s$ mesons are
$f_{B_s}=266$ MeV, $f_{B_s^*}=289$ MeV, and $f_{B_s^*}/f_{B_s} = 1.09$,
and these results show that the internal motion of quarks is less important in
the $B_s$ meson system compared to the the $D_s$ meson system, as expected.

Veseli and Dunietz \cite{VD} worked on the decay constants
of the $P$-wave scalar ($0^+_{1/2}$) and axial-vector ($1^+_{1/2}$) mesons
and derived
\begin{eqnarray}
F_{0^+_{1/2}}(p)&=&\Big[ {1 \over (m_q+E_q)} - {1 \over (m_{\bar{Q}}+E_{\bar{Q}})} \Big]
\, p\, R_{n1}(p)\ , \ \ \
\nonumber\\
F_{1^+_{1/2}}(p)&=& \Big[ {1 \over (m_q+E_q)} 
+ {1\over 3} {1 \over (m_{\bar{Q}}+E_{\bar{Q}})} \Big]
\, p\, R_{n1}(p) \ .
\label{ff7}
\end{eqnarray}
In the limit $m_{\bar{Q}}\to \infty$, both $F_{0^+_{1/2}}(p)$ and $F_{1^+_{1/2}}(p)$
become $pR_{n1}(p)/(m_{\bar{Q}}+E_{\bar{Q}})$ \cite{VD}.
However, for the $P$-wave $D_{sJ}$ mesons ($D_{sJ}^*$(2317) and $D_{sJ}$(2460))
the internal motion of quarks is even larger than that in the $S$-wave $D_s$ mesons,
and than the difference of $f_{0^+_{1/2}}$ and $f_{1^+_{1/2}}$ become much greater.
Using (\ref{ff5}) and (\ref{ff7}), Veseli and Dunietz \cite{VD} obtained the results:
$f_{0^+_{1/2}}=110$ MeV, $f_{1^+_{1/2}}=233$ MeV, and $f_{1^+_{1/2}}/f_{0^+_{1/2}} = 2.12$.
Their result for the ratio $f_{1^+_{1/2}}/f_{0^+_{1/2}}$
is very close to the value $f_{D_{s1}'}/f_{D_{s0}^*}\sim 2.26\pm 0.41$
presented in Table 1, and their results for $f_{0^+_{1/2}}$ and $f_{1^+_{1/2}}$ are
consistent with our results presented in Table 1:
\begin{equation}
|a_1|\, f_{D_{s0}^*}>74\pm 11\ {\rm MeV},\ \ \
|a_1|\, f_{D_{s1}'}>166\pm 20\ {\rm MeV},\ \ \
f_{D_{s1}'}/f_{D_{s0}^*}\sim 2.26\pm 0.41.
\label{fresults}
\end{equation}
Our results in (\ref{fresults})
also support that $D_{s0}^*$(2317) and $D_{s1}'$(2460) are
$j=1/2$ states instead of $j=3/2$ states, since Veseli and Dunietz \cite{VD}
obtained the values of decay constants
of the $D_{sJ}(1P,\ 1^+_{3/2})$ and $D_{sJ}(1D,\ 1^-_{3/2})$ states
as 87 and 45 MeV, respectively, which are much lower than
$|a_1|\, f_{D_{s1}'}>166\pm 20\ {\rm MeV}$ given in (\ref{fresults}).

We note that in the limit $m_{\bar{Q}}\to \infty$, $f_{0^-_{1/2}}$ and $f_{1^-_{1/2}}$
($f_{0^+_{1/2}}$ and $f_{1^+_{1/2}}$) become the same,
however, $f_{0^-_{1/2}}$ and $f_{0^+_{1/2}}$ ($f_{1^-_{1/2}}$ and $f_{1^+_{1/2}}$)
are different even in this heavy quark symmetry limit since $0^-_{1/2}$ and $1^-_{1/2}$
states are $S$-wave and $0^+_{1/2}$ and $1^+_{1/2}$ states are $P$-wave.
We can see this difference explicitly in (\ref{ff6}) and (\ref{ff7}).
Furthermore, the limit $m_{\bar{Q}}\to \infty$ does not corresponds to a good
approximation for the study of the $P$-wave $D_s$ meson system because of the large
internal motion of quarks inside the meson.
This property results in the fact that the decay constant of axial-vector meson
is about twice that of the scalar meson for the $P$-wave $D_s$ meson system.

\section{Conclusion}

The resonances $D_{sJ}^*$(2317) and $D_{sJ}$(2460) which is considered to be the
$(0^+,1^+)$ doublet composed of charm and strange quarks have been discovered recently.
Bell \cite{belle03} and Babar \cite{babar04} measured the branching ratios of
the exclusive modes
$B\to D D_{sJ}^*(2317)[D_s^+\pi^0]$,
$B\to D D_{sJ}(2460)[D_s^{*+}\pi^0]$
and $B\to D D_{sJ}(2460)[D_s^{+}\gamma ]$.
{}From these experimental data we extracted the lower bounds of the decay constants of
$D_{sJ}^*$(2317) and $D_{sJ}$(2460) by the method of Rosner which is based on
the factorization hypothesis.
Our result shows that the decay constant of $D_{sJ}$(2460) is about twice 
that of $D_{sJ}^*$(2317) on the contrary to the naive expectation of the heavy quark
symmetry which gives their equality.
We showed that this big deviation originates from the large internal motion of quarks
inside these $P$-wave states and that our result is in good accord with the relativistic
quark model calculation.
This result indicates that we can not apply the heavy quark symmetry to
$D_{sJ}^*$(2317) and $D_{sJ}$(2460).
For example, this result shows that the assumption of the heavy quark symmetry to
these states which was considered in \cite{DattaOD,ChenLi} is not valid.

Our results for the decay constants are given by
$|a_1|\, f_{D_{s0}^*}>74\pm 11\ {\rm MeV}$ and $|a_1|\, f_{D_{s1}'}>166\pm 20\ {\rm MeV}$,
where $|a_1|\sim 1$.
These results are consistent with the results of Veseli and Dunietz \cite{VD}
given by $f_{0^+_{1/2}}=110$ MeV, $f_{1^+_{1/2}}=233$ MeV, which were obtained from the
relativistic quark model calculation.
This fact is a good evidence that $D_{sJ}^*$(2317) and $D_{sJ}$(2460) are 
states with $j=1/2$ of the light degree of freedom, but not with $j=3/2$,
since the decay constants of the $D_{sJ}(1P,\ 1^+_{3/2})$ and $D_{sJ}(1D,\ 1^-_{3/2})$
states are much smaller than $166\pm 20\ {\rm MeV}$ which
is our result for the lower bound of $f_{D_{s1}'}$;
in the $m_{\bar{Q}}\to \infty$
the decay constants of the $D_{sJ}(1P,\ 1^+_{3/2})$ and $D_{sJ}(1D,\ 1^-_{3/2})$ states
become zero and the results from the relativistic quark model calculation by
Veseli and Dunietz \cite{VD} are given by 87 and 45 MeV, respectively.
When we use the results of Veseli and Dunietz \cite{VD} for the decay constants of
the $D_{sJ}(1P,\ 1^+_{1/2})$, $D_{sJ}(1P,\ 1^+_{3/2})$ and $D_{sJ}(1D,\ 1^-_{3/2})$ states,
we predict the ratio of the branching ratios,
${\cal B}(B\to D D_{sJ}(1P,\ 1^+_{1/2}))$ : ${\cal B}(B\to D D_{sJ}(1P,\ 1^+_{3/2}))$ :
${\cal B}(B\to D D_{sJ}(1D,\ 1^-_{3/2}))$ $\sim$ 1 : 0.14 : 0.04.
Therefore, it is clear that ${\cal B}(B\to D D_{sJ}(2460))$ measured by Belle and Babar
are consistent with $D_{sJ}(2460)$ being the $1^+_{1/2}$ state, but inconsistent with
being the $1^+_{3/2}$ or $1^-_{3/2}$ state.


\section*{Acknowledgments}
One of the authers (D.S.H.) wishes to thank Kazuo Abe, Vera Luth, and
Helmut Vogel for the helpful discussions.
This work was supported in part by the International Cooperation Program of
the KISTEP (Korea Institute of Science \& Technology Evaluation and Planning).

\end{document}